
\input phyzzx.tex
%
%
\def\tytul#1{}         
\def\neqn#1{\eqn#1}    
\def\E#1{#1}           
\def\name#1{}          
\def\Rc#1{{[#1]}}
\def\zb{{\bar z}}
\def\yb{{\bar y}}
\def\Db{{\bar D}}
\def\rhob{{\bar\rho}}
\def\dely{\partial_y}
\def\delz{\partial_z}
\def\delyb{\partial_{\bar y}}
\def\delzb{\partial_{\bar z}}
\def\delmu{\partial_\mu}
\def\delmub{\partial_{\bar \mu}}
\def\delnu{\partial_\nu}
\def\delnub{\partial_{\bar \nu}}
\def\delal{\partial_{\alpha}}
\def\delbe{\partial_{\beta}}
\def\deli{{\partial^i}}
\def\delj{{\partial^j}}
\def\dyb{d\bar y}
\def\dzb{d\bar z}
\def\Nb{{\bar N}}
\def\nb{{\bar n}}
\def\mub{{\bar\mu}}
\def\nub{{\bar\nu}}
\def\gMMb{ {g^{\mu\mub}} }
\def\xmu{{x^{\mu}}}
\def\xmub{{x^{\bar\mu}}}
\def\inv{^{-1}}
\def\dy{dy}
\def\dz{dz}
\def\pd#1#2{ {\partial #1 \over \partial #2} }
\def\R{{\bf R}}
\def\C{{\bf C}}
\def\hc{{{}^{\dag}}}
\Ref\OoguriVi{H. Ooguri and C. Vafa,
          \journal Mod. Phys. lett.&A5 (90) 1389.
           \tytul{Selfduality and N=2 string magic}\name{OoguriVi}}
\Ref\OoguriVii{H. Ooguri and C. Vafa,
          \journal Nucl. Phys. &B369 (91) 469.
           \tytul{Geometry of N=2 strings}\name{OoguriVii}}
\Ref\Ademello{M. Ademello et al
          \journal Nucl. Phys. &B111 (76) 77.
          \tytul{Dual strings with U(1) colour symmetry}\name{Ademello}}
\Ref\MathurMi{S.D. Mathur and S. Mukhi
          \journal Phys. Rev. &D36 (87) 465.
          \tytul{Becchi-Rouet-Stora quantization of twisted extended
                      fermionic strings}\name{MathurMi}}
\Ref\MathurMii{ S.D. Mathur and S. Mukhi
          \journal Nucl. Phys. &B302 (88) 130.
          \tytul{The N=2 fermionic string: Path integral, spin
                    structures and supermoduli on the torus}\name{MathurMii}}
\Ref\GreenSW{M.B. Green, J.H. Schwarz, and E. Witten,  Superstring Theory,
              (Cambridge University press, Cambridge, 1986) Vol I, p 237.
              \name{GreenSW}}
\Ref\deVega{H.J. de Vega
           \journal Commun. Math. Phys. &116 (1988) 659.
           \tytul{Non-linear multi-plane wave solutions of self-dual
                               Yang-Mills theory}\name{deVega}}
\Ref\MasonS{ L.J. Mason and G.A.J. Sparling
          \journal Phys. Lett. &A137 (89) 29.
          \tytul{Nonlinear Schr\"odinger and Korteweg- de Vries are
                    reductions of self dual Yang-Mills}\name{MasonS}}
\Ref\Wardi{R.S. Ward, ``Multi-dimensional integrable systems'' in
            `Field theory, quantum gravity and strings' Vol 2
               eds H. de Vega and N. Sanchez (1987).\name{Wardi}}
\Ref\Brihayeetal{Y. Brihaye, D.B. Fairlie, J. Nuyts and R.G. Yates
          \journal J. Math. Phys. &19 (78) 2528
          \tytul{Properties of the self dual equations for an SU(N) gauge
           theory}\name{Brihayeetal}}
\Ref\Pohlmeyer{K. Pohlmeyer
          \journal Commun. Math. Phys. &72 (80) 37.
          \tytul{On the lagrangian theory of anti self dual fields in
                   four dimensional Euclidean space}\name{Pohlmeyer}}
\Ref\Chauetal{L.-L. Chau, ``Geometrical integrability and equations of motion
          in physics: A unifying view'', in `Integrable Systems'
          edited by X.C. Song.
          \name{Chauetal}}
\Ref\BoniniGI{ M. Bonini, E. Gava and R. Iengo
          \journal Mod. Phys. Lett. &A6 (91) 795.
          \tytul{Amplitudes in the N=2 string}\name{BoniniGI}}
\Ref\Sokachev{S. Kalitzin and E. Sokatchev
          \journal Phys. Lett. &B262 (91) 444.
          \tytul{An action principle for selfdual Yang-Mills and Einstein
           equations}\name{Sokachev}}
\Ref\Parki{Q-Han Park
          \journal Phys. Lett. &B238 (90) 287.
          \tytul{Self-dual gravity as the large-N limit of the 2D
                    non-linear sigma model}\name{Parki}}
\Ref\Witteni{ E. Witten
         \journal Nucl. Phys. &B311 (88) 46.
         \tytul{2+1  gravity as an exactly soluble system}\name{Witteni}}
\Ref\Yangi{C.N. Yang
          \journal Phys. Rev. Lett. &38 (77) 1377.
          \tytul{Condition for the self-duality for SU(2) gauge fields
                       on Euclidean four-dimensional space.}\name{Yangi}}
\Ref\Corriganetal{ E.F. Corrigan, D.B. Fairlie, R.G. Yates and P. Goddard
          \journal Commun. Math. Phys. &58 (1978) 223
          \tytul{The construction  of self-dual solutions to SU(2) gauge
          theory}\name{Corriganetal}}
\Ref\Green{M.B. Green
           \journal Nucl. Phys. &B293 (87) 593.
           \tytul{World sheets for world sheets}\name{Green}}
\nopubblock
\titlepage
\line{\hfil ETH-TH/91-35}
\line{\hfil August 1991}
\title{On N=2 strings and classical scattering solutions of self-dual
Yang-Mills in (2,2) spacetime}
\author{Andrew Parkes}
\address{Theoretische Physik,\break
ETH-H\"onggerberg,\break
8093 Z\"urich,\break
Switzerland.}
\vfil
\abstract
Ooguri and Vafa have shown that the open N=2 string corresponds to
self-dual Yang-Mills (SDYM) and also that, in perturbation theory, it has has
a vanishing four particle scattering amplitude.  We discuss how the dynamics
of the three particle scattering implies that on shell states can only scatter
if their momenta lie in the same self-dual plane and then investigate
classical SDYM with the aim of comparing exact solutions with the tree level
perturbation theory predictions.  In particular for the gauge group SL(2,C)
with a plane wave Hirota ansatz SDYM reduces to a complicated set of algebraic
relations due to de Vega.  Here we solve these conditions and the solutions
are shown to correspond to collisions of plane wave kinks.  The main result is
that for a class of kinks the resulting phase shifts are non-zero, the
solution as a whole is not pure gauge and so the scattering seems non-trivial.
However the stress energy and Lagrangian density are confined to string like
regions in the space time and in particular are zero for the incoming/outgoing
kinks so the solution does not correspond to physical four point scattering.
\endpage
\chapter{Introduction}

Recently Ooguri and Vafa \Rc{\OoguriVi,\OoguriVii}\ have revived interest in
string theories with two local worldsheet supersymmetries.  Such string
theories were investigated early in the days of dual models \Rc{\Ademello}, and
were found to have a critical dimension of two.  However only comparatively
recently was it realised that these are two complex dimensions \Rc{\MathurMi}.
Thus, they naturally live in a real space-time with a (2,2) signature.  A
string in (1,1) real dimensions cannot have any transverse oscillations and so
it is entirely natural that it does not give rise to the usual infinite tower
of massive states.  For a string in (2,2) dimensions it might be expected that
transverse modes do occur, but for the N=2 string these turn out to
correspond to local N=2 supersymmetry transformations and so do not give rise
to physical particles, and so the spectrum contains only a finite number of
particles.  It turns out that they are all massless, bosonic and scalar (see
\Rc{\MathurMi,\MathurMii, \GreenSW,\OoguriVi,\OoguriVii}\ and references
therein for more details).

Ooguri and Vafa investigated the string scattering amplitudes in order to
identify the field theory corresponding to the particles in the spectrum. From
the closed N=2 string three particle scattering they found that the
scalar can be interpreted simply as the K\"ahler potential of a Ricci flat
complex manifold.  Such manifolds can equivalently be described as self-dual
gravity (SDG).  This is in line with the fact that if one tries to couple the
world sheet action to the geometry of the space-time then the only way to do
this and preserve the N=2 supersymmetry is to couple it only via the K\"ahler
potential. It was also shown that open strings with Chan-Paton group factors
attached at their edges gave rise to self-dual Yang-Mills (SDYM) theories.

The four particle string amplitude usually consists of products of gamma
functions and so there will be a infinite number of values for the Mandelstam
s, t and u variables that will will lead to a pole in the scattering
amplitude.  For N=0 and N=1 strings these poles just correspond to the massive
particles in the spectrum. However for N=2 strings there is no such infinite
tower of massive states and so there is a potential loss of unitarity. The
solution is simply that the kinematic factor in front of the four particle
amplitude vanishes whenever the external particles are on shell.  This fact
relies on a special identity concerning any three null vectors in a (2,2)
space-time \Rc{\OoguriVii, but see also \deVega}.  Similarly it is believed
that all the higher point connected on-shell amplitudes also vanish.  This
gives an almost trivial solution to the problem of finding amplitudes that are
dual with respect to the Mandelstam variables.  This is certainly a very
remarkable property for a quantum field theory to have and this paper is
inspired by the hope of shedding some light on this property.

Since there are no higher mass states to integrate out it follows that the
field theory (SDYM or SDG) from the N=2 string theory is not just a low energy
theory and so at tree level the theories should be equivalent.  When loops are
included then they differ because the scalars act like Liouville modes with
non-standard rules for path integrals and also because modular invariance
changes regions of integration in the loops.  Even at tree level the string
approach has the advantage that the perturbation theory combines all connected
field theory Feynman diagram into a single string diagram. However given the
difficulty of a non-perturbative approach to string theory it is much easier
to deal with exact solutions for the field theory.  Hence in this paper we
look again at classical SDYM in light of the results from the string theory.

It is well known that classical SDYM is an integrable system.  For the quantum
version of integrable systems in (1,1) dimensions we would usually expect to
have an infinite number of conserved charges that prohibit any particle
production and also for the S matrix to factorise into products of S matrices
for $2\to2$ scattering.  However, the string theory seems to predict that SDYM
in (2,2) has almost the exactly opposite properties; that there is no
(2$\to$2) scattering and that there is particle production from the (1$\to$2)
S-matrix contribution.  Of course, we will ignore the infrared problems that
would occur in a careful discussion of scattering of massless states.

We observe that the kinematical properties of the string theory results of
\Rc{\OoguriVi}\ suggest that the system exhibits self-dual null plane
decoupling, meaning that physical particles only interact, at the S-matrix
level, if their momenta lie in the same selfdual null plane.  Specifically for
a massless particle with momentum $k$ in coordinates $(yz\yb\zb)$ we define
$\omega=k_z/k_\yb$ and this parametrises the self-dual null plane in which it
lies.  Particles with different $\omega$ values would not see each other at
the S-matrix level, suggesting that the theory is not fully four dimensional.
If valid in all cases then this would presumably have strong effects on the
allowed S-matrices in these theories.  This is rather reminiscent of the
decoupling of left and right movers in two dimensional conformal field theory.
Thus, it is of interest to investigate whether the stringy results of
\Rc{\OoguriVi}\ are of general application.

One obvious proviso is that the string theory predictions are only a
perturbative result and so could be misleading. Furthermore, even the
perturbation theory prediction of the lack of connected $2\to2$ scattering has
the potential problem that the calculation only makes sense when it is
non-singular, that is, when $stu\neq0$.  Hence the calculation does not
directly rule out contributions of the form $\delta(s)$ or similar.  Normally
we would exclude such a possibility by analyticity of the S-matrix, but it is
not so clear that this is also true for integrable theories or for (2,2)
dimensions.

In particular for an integrable theory we would expect there to be solutions
where the scattering does not induce a change of momentum but only a phase
shift.  Thus, if the momenta of the external particles are $k_i$ with
$i=1,2,3,4$ and $k_i^2=0$, $\Sigma_i k_i=0$ then we will be particularly
interested in the case for which $k_3=-k_1$ and $k_4=-k_2$. This has $(k_1 +
k_3)^2=0$ and so corresponds to a singular set of Feynman diagrams (or a
string theory calculation at the boundary of moduli space where two of the
vertex insertions coincide on the world sheet) and so these may not be
reliable.  Classically we expect this to correspond to the case in which the
fields only depend on two of the four coordinates.  For example it is known
that under certain conditions SDYM is equivalent to the two-dimensional KdV
equation \Rc{\MasonS}, and this certainly has solutions that display
non-trivial soliton scattering.  The relevance of such ``phase shift only''
solutions is that $k_1$ and $k_2$ will generally not lie in the same self-dual
null plane.  If such scattering is truly relevant to SDYM then the decoupling
of these planes would not occur, thus contradicting the perturbative results.
Hence in this paper we take a (very restricted) look for classical solutions
of SDYM that correspond to $2\to2$ scattering or to $1\to2$ scattering and
attempt to clarify the situation with respect to classical scattering.

 We emphasise that when talking of the triviality, or otherwise, of the
``classical S-matrix'' it is important to specify the type of waves that are
colliding and whether they are to be considered as physical.  As pointed out
in \Rc{\deVega}, based on the example of the KP system, it can well be that
finite size packets do not scatter but that plane waves might still suffer a
non-trivial phase shift. In an integrable theory the infinite number of
conserved charges might seem to restrict scattering but for a plane wave the
infinite extent of the wave front could lead to infinite values for these
charges thus invalidating the conservation laws.

Dimensional reduction of SDYM is of interest in its own right because many of
the usual integrable systems can be generated in this fashion (for example
\Rc{\MasonS,\Wardi}).  One can hope that SDYM is then a master integrable
system for many others.  However the emphasis here is not to generate new
reductions but to see what the reductions say about SDYM itself.

A slightly discordant note on the N=2 string is the redundancy of having a
plethora of fermions on the world-sheet, and yet the theory contains no
spacetime fermions.  Also, since we have a (1,1) worldsheet embedded into a
(2,2) space-time then the transverse modes must be pure gauge in order to
explain the lack of higher mass states.  Hence we need to identify the
nilpotent translations obtained by a supersymmetry transformation on the
worldsheet with a non-nilpotent translation in the space-time coordinates and
it seems unnecessary to have to face this problem.  This reinforces the
suggestion in \Rc{\OoguriVii}) that the theory has a formulation in terms of
some purely bosonic extended object.

In chapter 2 we discuss the peculiarities of kinematics in a (2,2) space-time.
In chapter 3 we review the J-formulation of SDYM, in which the Yang-Mills
fields are written as derivatives of a pre-potential\Rc{\Brihayeetal,
\Pohlmeyer, \Chauetal}.  The pre-potential J is a scalar and seems to be the
natural outcome of the N=2 string.  The J-formulation is also a natural
generalisation of the WZNW model in two dimensions.  For the reason given
above we are particularly interested in the case in which J depends on only
two coordinates $\eta_1,\eta_2$ for which it is trivial to see that the system
reduces to chiral models with Wess-Zumino term in (1,1) dimensions.  In
certain cases the system has obvious exact generic  solutions and we  relate
these to the string predictions for  three particle scattering.

In chapter 4 we consider the special case in which the gauge group is SL(2,C).
This group has the advantage that a Gauss decomposition for the group element
leads to Yangs equations which are derivable from an action.  Using the tree
level Feynman rules we verify that the connected four point contributions sum
to zero for generic external null momenta.  Also Yangs equations are
homogeneous in the fields and de Vega exploited this to search for scattering
solutions.  With a fractional Hirota ansatz he reduced Yangs equations to a
large set of algebraic equations, and was able to produce some scattering
solutions.  Here we show that with a good change of variables these equations
 have a reasonably simple general solution and so we obtain a wide class
of classical scattering solutions.  The solutions we obtain correspond to
kink-kink collisions.  They are non-trivial in the sense that the collision
does indeed generate a phase shift for the kinks, but in order to get the
phase shift we are forced to impose conditions on the asymptotic states which
mean precisely that they have have zero value of the relevant stress energy
tensor.  In chapter 5 we conclude with a discussion of our results and their
implications along with a few extra observations.

\chapter{Kinematics in (2,2) space-time}

Consider a  complex four dimensional space with coordinates
$x^\alpha=(x^\mu,x^\mub)$ where $x^\mu=(y,z)$, $x^\mub=(\yb,\zb)$, and with a
complex metric
$$ ds^2 = g_{\alpha\beta} dx^\alpha dx^\beta = 2 g_{\mu\mub} dx^\mu dx^\mub =
2 (\dy \dyb -  \dz \dzb)
\neqn\metricdefs$$
There are two simple ways to recover a real (2,2) space-time.  If we impose
$(x^\mu)^* = x^\mub$ then we still have complex coordinates and so call this
a $C^{1,1}$ slice; the metric has a manifest holomorphic U(1,1) symmetry.
Alternatively we can impose that $x^\mu$ and $x^\mub$ are independent real
coordinates, we call this a $R^{2,2}$ slice.  For the sake of comparison with
a real (3,1) space time we could also consider an $\R^{3,1}$ slice in which
$\yb=y^*$ but $z\in\R$ and $\zb\in\R$.  Given momenta $k_\alpha =
(k_\mu,k_\mub) = (k_y,k_z,k_\yb,k_\zb)$, we use the non-symmetric scalar
product
$$ <p\vert k> := g^{\mu\mub} p_\mu k_\mub = p_y k_\yb - p_z k_\zb
\neqn\innerprod$$
The symmetric scalar product is $p\cdot k := g^{\alpha\beta}p_\alpha k_\beta =
<p\vert k> + <k \vert p> $.  We will later consider vectors $k_i$ and it is
convenient to define $k_{ij} := <k_i \vert k_j >$, $s_{ij} := k_{ij} + k_{ji}$
and $c_{ij} := k_{ij} - k_{ji}$.

It turns out that all the interactions in the string theories or in SDYM or
SDG are simple functions of the $k_{ij}$; even one-loop corrections do not
seem to affect this conclusion because the kinematic factors are similar to
the tree level \Rc{\BoniniGI}.  So, before proceeding to the specific case
of SDYM, we can first discuss some general kinematical properties of the
on-shell amplitudes.

For any 2x2 matrix M it is trivially true that
$$
M_\mu{}^{[\mub} M_\nu{}^{\nub} M_\sigma{}^{\bar\sigma]} = 0
\neqn\asymM$$
where the square brackets mean antisymmetrisation.
So if we take $M_\mu{}^\mub = g_{\mu\mub}$ and contract the free indices
with three arbitrary vectors $k_i$ we obtain a cubic identity in $k_{ij}$.
For the particular case that the vectors are null we obtain
$$
k_{12} k_{23} k_{31} + k_{21} k_{32} k_{13} =0
\neqn\dreinull$$
This ``three nulls identity'' is valid in the four complex dimensional space
and so is also valid for any restriction to a real four-dimensional space
time.  If the real slice has a (4,0) signature then there are no non-zero null
vectors and so \E\dreinull\ is empty.  If the real slice is (3,1) then this is
essentially the same as the identity found in \Rc{\deVega}\ and shown to be
relevant to the consistency of multisoliton scattering in SDYM with SL(2,C)
gauge group.  When the real slice has signature (2,2) then as shown in
\Rc{\OoguriVi}\ it is responsible for the vanishing of the kinematic factor
in the four particle on-shell string amplitude.

If we consider on-shell scattering of three massless particles with momenta
$k_i$, then conservation of momentum implies that the external momenta span a
null plane.  The bivector $k_{1\alpha} k_{2\beta} - k_{1\beta} k_{2\alpha}$
that defines the null plane can be either anti-self-dual or self-dual.  The
anti-self-dual case means that $k_i=(p_i,p_i\Omega,\bar p_i,\bar
p_i\bar\Omega)$ with $\Omega\bar\Omega=1$, whilst for a self-dual plane we
have $k_i=(p_i,\bar p_i \omega,\bar p_i, p_i \bar\omega)$ with
$\omega\bar\omega=1$.  However for the case of the anti-self-dual plane we see
that $k_{ij}=0$ and so there is no interaction in the theories under
consideration.  In other words, if we write the momenta of all particles in
the form $(p_i,\bar p_i \omega_i,\bar p_i, p_i \bar\omega_i)$ then particles
can only interact with each other via the on-shell three leg vertex if they
have the same $\omega_i$ value and so lie in the same self-dual null plane.
If we convert the momenta to operators on some field $\phi$ we get $(\delz -
\omega\delyb)\phi = (\delzb - \bar\omega\dely) \phi= 0$ and this corresponds
to what we shall call an ``$\omega$-ansatz''
$$ \phi = \phi( y + \bar\omega \zb, \yb + \omega z )
\neqn\omansatz$$
In a  $\R^{3,1}$ slice
there are no null planes but only null lines and so cannot deal with
scattering in this way.

The three nulls identity \E\dreinull\ gives rise to the
following identities, valid for any momenta $k_i$ $i=1,\ldots,4$ with
$k_{ii}=0$ and $\Sigma_i k_i=0$
$$
{k_{13} k_{42} \over s_{13}} + { k_{12} k_{43} \over s_{12} } = 0
\qquad
{k_{13} k_{24} \over s_{13}} + { k_{12} k_{34} \over s_{12} } + k_{14} = 0
\neqn\kidens$$
$$ {c_{13} c_{24} \over s_{13}} + { c_{12} c_{34} \over s_{12} } + s_{14} = 0
\neqn\cidens$$
and these will be used to compute the four particle scattering.

Finally consider the kinematics of the four massless particle on-shell
amplitude.  For this paragraph only, let us suppose that we have a real
diagonal metric $(-,+,+,\pm)$, and that $s_{12}\neq0$.  Then after rescaling
and rotations of momenta the generic situation is that $k_1=(1,1,0,0)$,
$k_2=(1,-1,0,0)$, $k_3=(-1,p_1,p_2,0)$ and $k_4=(-1,-p_1,-p_2,0)$ where $p_1^2
+ p_2^2=1$ and so $s_{12}=-4$, $s_{13}=2+2p_1$, and $s_{14} = 2 - 2 p_1$.  The
sign of $g_{33}$ is irrelevant and so we don't see the difference in this case
between a (2,2) and a (3,1) signature.  The point here is simply that in a
(2,2) signature there exist reasonable, non-parallel, sets of physical momenta
for the on-shell three and four particle amplitudes and so these amplitudes
can have direct physical significance.

\chapter{SDYM and the J formulation}

In order to set up SDYM in a complex space we take a complex gauge group
$G_C$, and take $F_{\alpha\beta}=[\nabla_\alpha,\nabla_\beta]$ with
$\nabla_\alpha =\partial_\alpha + A_\alpha$.  The (anti)-self-duality
condition is that $F_{\alpha\beta}= - {1\over2} \det (g_{\mu\mub})
\epsilon_{\alpha\beta\gamma\delta} F^{\gamma\delta}$ where
$\epsilon_{yz\yb\zb}=+1$.  In  the above metric this reduces to
$$ F_{\mu\nu} = F_{\mub\nub} =0 \qquad\qquad \gMMb F_{\mu\mub}=0
\neqn\sdcondits$$
Hence there exist independent group elements $D$ and $\Db$ such that
$$
A_\mu = D^{-1} \delmu D
\qquad\qquad
A_\mub = \Db^{-1}\delmub \Db
\neqn\AAb$$
Yang-Mills gauge transformations now correspond to $D \to D g$ and $\Db \to
\Db g$ with $g\in G_C$ and so the gauge invariant quantities live on $(G_C
\otimes G_C)/ G_C$. A natural representative for each orbit is the gauge
invariant quantity
$$
J := D \Db^{-1}
\neqn\defJ$$
and the self duality condition reduces to
$$
\delyb ( J\inv \dely J) - \delzb ( J\inv \delz J) =
\gMMb \delmub ( J^{-1} \delmu J) =0
\neqn\pohl$$
In particular this implies that $\gMMb \delmu \delmub (\ln \det J)=0$ so $\det
J$ is simply a free field and is of no dynamical interest.  In the process of
making the above gauge choices to go to the J formulation we have reduced the
manifest Lorentz invariance, although of course gauge invariant quantities
still must be fully Lorentz invariant.  However we do gain the semi-local $G_C
\otimes G_C$ symmetry
$$
J \to g_L(\xmub)J g_R(\xmu)
\neqn\pohlsyms$$
By definition $J$ is gauge invariant, however in order to recover any
Yang-Mills structures from it we need to split first into $D$ and $\Db$ and
the arbitrariness of this split is the Yang-Mills gauge symmetry. Thus, gauge
potentials and field strengths are gauge dependent functions of the gauge
invariant $J$.  The symmetry \E\pohlsyms\ clearly leaves the Yang-Mills
potentials invariant and so $J$ is not uniquely given by the field strengths.
Also, given the field strengths it is a non-local process to find a
representative $J$, however the $J$ formulation could be considered as just as
fundamental as the Yang-Mills formulation.

The classical integrabilty of the system is revealed by fact that \E\pohl\ is
simply the compatibility condition for the linear system $\epsilon^{\mu\nu} J
\delnu \Psi + \lambda \gMMb \delmub(J\Psi) \equiv 0$ for all $\mu$ and all
$\lambda$.  From this an infinite number of conservation laws can be derived
in the standard fashion (see \Rc{\Chauetal} and references therein).

Ultimately we need to impose reality conditions on the coordinates and if we
also wish to finish with a real gauge group $G_R$ then we need to impose
reality restrictions on $D$ and $\Db$ and so on $J$.  For the $\R^{2,2}$ slice
we can trivially take $D,\Db,J\in G_R$ so that ${A_\alpha}{}^{\dag}
=-A_\alpha$.  For the $\C^{1,1}$ case we need ${A_\mu}{}^{\dag}=-A_\mub$ and
so must take $\Db = (D\hc)^{-1}$ and this reality condition is preserved by
gauge transformations satisfying $g\hc g =1$; hence the gauge invariants in
this case live on $G_C / (G_C)_U$ where $(G_C)_U$ is the subgroup of unitary
elements of $G_C$. In this case $J=D D\hc$ becomes a positive hermitian
element of $G_C$. The fundamental example is to take $G_R$=U(N) and $G_C$ to
be its complexification GL(N,C), and then J is a positive hermitian NxN matrix.
Even though $\C^{1,1}$ and $\R^{2,2}$ are of course trivially related by
coordinate changes the corresponding J fields are not the same, or even in
the same space in general.

The equation of motion \E\pohl\ shows that the $J$ formulation is a
dimensional generalisation or complexification of the
Wess-Zumino-Novikov-Witten (WZNW) model in (1,1) space-time dimensions and so
one cannot expect directly to have a manifestly local and group invariant
action.  An action is feasible if we pick a particular parametrisation for the
group (see the next chapter) or if one includes extra coordinates
\Rc{\Sokachev}.  However without an action we can obtain some Feynman rules
from \E\pohl.  Presumably the properties of the N=2 string are true in any
background that solves the equations of motion, but here we will only consider
expansions around trivial flat backgrounds.  Expanding about $J=1$ gives a
$1/p^2$ propagator and vertices with arbitrary numbers of external legs but
which are all proportional to $c_{12}$

If one now looks at four particle on-shell tree level scattering then the
three diagrams have intermediate (squared) momenta $s_{12}$,$s_{13}$ and
$s_{23}$ and so we would normally expect their differing momentum dependencies
to be forbid any cancellation.  However, due to the identity \E\cidens\ the
Feynman diagram with an apparent $s_{13}$ pole can be changed into an apparent
$s_{12}$ channel pole and so it is now reasonable that the three diagrams
combine into one term linear in $k_{ij}$ which can than cancel the
contribution from the four point vertex.  It is then straightforward to check
that the connected four point tree level diagrams do indeed sum to zero when
the external legs are all on shell.  This of course matches the expectation
from the N=2 string theory \Rc{\OoguriVi,\OoguriVii}; but it is interesting to
note that we do not need to use the reality conditions at any time and so this
is actually true in $\C^4$.

For reasons given in the introduction we now dimensionally reduce SDYM, in the
same fashion as \Rc{\deVega}, by considering generalised null plane waves.
$$ J = J[\eta_1,\eta_2]
\neqn\genplane$$
where $\eta_i=k_{i\alpha}x^\alpha=k_{i\mu}x^\mu+k_{i\mub}x^\mub$ and the $k_i$
are linearly independent and null.  Using $\deli = \partial/\partial \eta_i$
and $c_{ij}=\epsilon_{ij} c_{12}$ then the self-duality condition
\E\pohl\ gives
$$
s_{12} [\partial^1 (J\inv \partial^2 J ) + \partial^2 (J\inv \partial^1 J )] +
c_{12} \epsilon_{ij} \delj (J\inv \deli J ) =0
\neqn\redpohl$$
The case of $k_2=0$ is trivial: if $k_1^2=0$ then $J[\eta]$ is arbitrary.
Hence any null plane wave solves both the full non-linear equations and their
linearised versions which is unusual for a non-linear system.  If we were to
attempt a classical version of S-matrix theory perturbation theory then we
would usually want to switch off the interaction at infinity for the in and
out states.  This switching off of the interaction is not very pleasing, but
does not seem to be necessary for SDYM (or for SDG).

If neither $k_i$ vanishes then \E\redpohl\ just the chiral model with
Wess-Zumino term (CMWZ) model (1,1) space-time.  To us the most interesting
case is when the momenta lie in the same null plane and so can correspond to
on-shell three particle scattering.  This means that $s_{12}=0$ and we see
that the corresponding reduction of SDYM is not an evolution equation.  This
is related to the previous observation that any null wave is a solution.  To
see this heuristically recall that an evolution equation will have an initial
surface on which the field and some finite number of derivatives are
specified, but one can (at least locally) pick a direction perpendicular to
this and since this direction is also null the solution can depend arbitrarily
on it and hence cannot be specified by a finite number of initial values.
More specifically we have two cases: If the $k_i$ lie in an anti-self-dual
plane then $c_{12}=0$ and so $J$ is totally arbitrary compared to a string
prediction that there is no scattering.  Whilst if the $k_i$ form a self-dual
plane then we have
$$
\epsilon_{ij} \deli (J\inv \delj J ) =0
\neqn\pureWZ$$
which corresponds to a two dimensional (topological) theory with only the WZ
term and no kinetic term.  The generic solution of this ``pure WZ'' system is
that $J=J[f(x^\alpha)]$ where $J\in G_C$ and $f \in \R$ are arbitrary.  The
fact that we do not obtain evolution equations for the cases which correspond
to the kinematics of the string or SDYM derived three point vertex suggests
that we need to be very careful when talking of the implications for
scattering in the theory.  In particular it is obviously possible, but not
very meaningful, to write down functions $J$ and $f$ that look like any number
of solitons ``scattering'' into any number of others; scattering of two into
one is allowed but not at all special.

In light of the above relation between pure WZ models and self-dual systems we
briefly mention some work of Park \Rc{\Parki}.  With some rearrangement we can
rewrite \Rc{\Parki} in the notation of this paper and in a more symmetric
fashion. Consider a tensor field with components $G_{\mu\mub}(x^\alpha)$ and
$\bar G_{\mub\mu}(x^\alpha)$ and define the vector fields
$$
A_\mu := G_{\mu\mub} \epsilon^{\mub\nub} \pd{}{x^\nub}     \qquad\qquad
\bar A_\mub := \bar G_{\mub\mu} \epsilon^{\mu\nu} \pd{}{x^\nu}
\neqn\vecsxy$$
So each $A_\mu$ generates diffeomorphisms (not necessarily area preserving) on
the self-dual null plane $x^\mu=$constant. Similarly each $A_\mub$ gives
diffeomorphisms on the self-dual null plane $x^\mub$= constant.
Introduce  the derivative $\nabla_\alpha =
\partial_\alpha + e A_\alpha$ where e is a coupling constant and the  field
strength operator $F_{\alpha\beta}=[\nabla_\alpha , \nabla_\beta]$.  We then
enforce the operator condition that
$$
F_{\mu\nu}=F_{\mub\nub}=0 \qquad\quad \hbox{for all} \quad e
\neqn\Parkcondits$$
The O($e$) term gives
$$
\epsilon^{\mub\nub} \delmub G_{\rho\nub}=0  \qquad
\epsilon^{\mu\nu} \delmu \bar G_{\rhob\nu}=0  \qquad   \forall \rho, \rhob
\neqn\orde$$
with solution that $G_{\mu\mub} = \delmub F_\mu$ and $\bar G_{\mub\mu} =
\delmu \bar F_\mub$ for some functions $F$, $\bar F$. In particular this
forces the diffeomorphisms to be area preserving on the corresponding planes
(here our emphasis differs form \Rc{\Parki}).  To match Park we now force
$G_{\mu\mub}=\bar G_{\mub\mu}$ so that we can regard it as a metric on the
space-time (this stage seems ad hoc and perhaps instead we should allow
coupling to an antisymmetric tensor).  Then \E\orde\ gives
$G_{\mu\mub}=\delmu\delmub\Omega$ and so this metric is K\"ahler.  With these
conditions on $G_{\mu\mub}$ the O($e^2$) terms of \E\Parkcondits\ have two
consequences.  Firstly
$$
G^{\mub\mu} F_{\mu\mub} = 0 \qquad\quad \hbox{for all} \quad e
\neqn\sdalle$$
so the field strength F is ``anti-self-dual for all e'' with respect to the
metric $G_{\mu\mub}$.  Secondly $\det G_{\mu\mub}$ is forced to be a constant
and so $G_{\mu\mub}$ defines a Ricci flat K\"ahler metric on the spacetime,
meaning that we have SDG.  Hence SDG in (2,2) space-time is almost SDYM with a
gauge group which consists of diffeomorphisms acting on self-dual null planes
embedded in the spacetime, but with the extra conditions that the self-duality
holds true even under constant rescalings of the gauge fields and that the
resulting metric is symmetric.

The relation to pure WZ models is simply that $F_{\mu\nu}=0$ for all e, has
the solution $ A_\mu = D^{-1} \delmu D$ with $\epsilon^{\mu\nu} \delmu
(D^{-1} \delnu D) = 0$ which is a pure WZ model on the $x^\mu$-plane for a
fixed point in $x^\mub$ (and similarly for the barred quantities).

As discussed in \Rc{\OoguriVii}\ the N=2 string theories seem to be almost
topological theories. It is also well known that the Yang-Mills stress
energy tensor vanishes for SDYM and this would be a sign that the theory is
topological were it not for the fact that the self-duality condition itself
uses a metric.  So it is interesting to note that the system discussed above
does not any reference to a metric on the space-time but only assumes a
complex structure; that is, we only needed the conditions $G_{\mu\mub}=\bar
G_{\mub\mu}$ and the tensors $\epsilon^{\mu\nu}$ and $\epsilon^{\mub\nub}$ for
\E\Parkcondits.  Instead the metric arises from the parameters of a gauge
transformation.  Also, it is reminiscent of Wittens work on (2,1) gravity
\Rc{\Witteni}\ that the usual Yang-Mills perturbation theory around
$A_\mu=\bar A_\mub=0$ corresponds to expanding about $G_{\mu\mub}=0$ and not
around the usual flat metric.  (However, since the conditions are to be
enforced for all $e$ the standard perturbation theory will not apply.)

For completeness, we note that for SDG expanding around a flat metric gives
the Plebanski equation
$$
\gMMb \delmu\delmub \phi +
\epsilon^{\mu\nu} \epsilon^{\mub\nub} (\delmu \delmub \phi) \delnu \delnub \phi
= 0
\neqn\plebinphi$$
If we then dimensionally reduce as we did for SDYM by imposing
$\phi=\phi(\eta_1,\eta_2)$ with $k_{11}=k_{22}=0$ we get
$$
s_{12} \partial_1 \partial_2 \phi + k_{12} k_{21} [
(\partial_1^2 \phi) \partial_2^2 \phi - (\partial_1 \partial_2 \phi)^2] =0
\neqn\redpleb$$
Then $s_{12}=0$, $k_{12}\neq0$ forces
$$
(\partial_1^2 \phi) \partial_2^2 \phi - (\partial_1 \partial_2 \phi)^2 =0
\neqn\plebom$$
which  has the generic solution that either
$\phi=\phi(a \eta_1 + b \eta_2)$ where a and b are arbitrary constants, or
that $\phi =(\eta_2 - \tilde \eta_2)
 \Phi[(\eta_1 - \tilde \eta_1)/(\eta_2 - \tilde \eta_2)] + b$ where
$\tilde\eta_i$,$b$ are arbitrary constants and $\Phi$ is an arbitrary function.
Again these do not behave as scattering solutions.

\chapter{SDYM with gauge group SL(2,C)}

This has been extensively studied previously; the motivation was either to
impose reality conditions and so  obtain self-dual SU(2) solutions
\Rc{\Yangi,\Pohlmeyer},  or to exploit the fact that it is a complex group
and so can have self-dual solutions in a (3,1) metric \Rc{\deVega}.  We
study it simply because it is the simplest non-trivial case.  The most
straightforward approach to SL(2,C) is based on the Gauss decomposition
$$\eqalign{
J &= \exp g \rho \pmatrix{ 0 & 0 \cr 1 & 0 \cr}
  \exp g S \pmatrix{ 1 & 0 \cr 0 & -1 \cr}
  \exp g \rhob \pmatrix{ 0 & 1 \cr 0 & 0 \cr}  \cr
  &= \pmatrix{ 1 & 0 \cr g \rho & 1 \cr}
    \pmatrix{ e^{g S} & 0 \cr  0 & e^{-g S} }
    \pmatrix{  1 & g \rhob  \cr 0 & 1 }
\cr}\neqn\Gauss$$
where $S$, $\rho$, and $\rhob$ are independent complex fields.
The equations of motion from \E\pohl\ are then
$$\eqalign{
\gMMb( - \delmu \delmub S + g e^{2gS} \delmu \rho \delmub \rhob )& =0 \cr
\gMMb \delmu ( e^{2gS} \delmub \rhob ) =
\gMMb \delmub ( e^{2gS} \delmu \rho ) &= 0
\cr}\neqn\yangsinS$$
which follow from the Lagrangian \Rc{\Brihayeetal}
$$ L = \gMMb(\delmu S \delmub S + e^{2gS} \delmu\rho \delmub\rhob)
\neqn\newpohlaction$$
To put this in context we note that dimensional reduction, as in the previous
chapter, would give a manifestly local action for the WZNW model in two
dimensions, but of course manifest group invariance has been lost.  From the
Lagrangian we obtain a stress-energy tensor
$$\eqalign{
T_{\mu\nu} &= \delmu S \delnu S + e^{2gS}  \delmu\rho \delnu\rhob \qquad
T_{\mub\nub} = \delmub S \delnub S + e^{2gS}  \delmub\rho \delnub\rhob \cr
T_{\mub\mu} &=
  T_{\mu\mub} = \delmu S \delmub S + e^{2gS}  \delmu\rho \delmub\rhob \cr}
\neqn\TdV$$
the equations of motion \E\yangsinS\ imply that $\partial^\alpha
T_{\alpha\beta}=0$.  We see that $T_{\alpha\beta}$ is symmetric, but not
covariant because of the way in which $\rho$ and $\rhob$ appear.  This T is of
course different from the stress tensor from the Yang-Mills action.

If we only wish to work at tree level then we can ignore any ghosts that arise
from the gauge fixing of the action from \E\newpohlaction, and so it is easy
to obtain the Feynman rules.  In particular, after dropping total derivatives,
the quadratic part of the Lagrangian is simply $L_0 = - {1\over2}
g^{\alpha\beta} (\delal S \delbe S + \delal \rho \delbe \rhob)$ giving $1/p^2$
propagators. The vertex with a $\rho$ of momentum $k_1$, a $\rhob$ of momentum
$k_2$ and n legs of $S$ comes with a factor $(2g)^n k_{12}$.
It is now straightforward to sum the connected tree-level diagrams with four
on-shell external legs.
Then
$$\eqalign{
<\rho(k_1)\rhob(k_2)\rhob(k_3)\rho(k_4)> & \;\propto\;
 (2g)^2 \left(
{k_{13} k_{42} \over s_{13}} + { k_{12} k_{43} \over s_{12} } \right)   \cr
<\rho(k_1)S(k_2)S(k_3)\rhob(k_4)> & \;\propto\; (2g)^2 \left(
{k_{13} k_{24} \over s_{13}} + { k_{12} k_{34} \over s_{12} } + k_{14} \right)
\cr}\neqn\fourpoint$$
Putting $k_{ii}=0$ and $\Sigma_i k_i=0$ and using \E\kidens\ shows that in any
case for which the calculation is non-singular then the answer is zero.  It is
interesting that this happens despite the fact that the map from the variables
$\lambda_a$ of the previous chapter to $S,\rho,\rhob$ is very non-linear.
Non-linear changes of field variables in this case do not affect the
conclusion that the connected four particle S matrix is zero.  We also note
that $\rho$ and $\rhob$ appear only quadratically in the action so that in
principle we could integrate them out, however this does not seem to be
directly useful.

We now want to exhibit a class of exact solutions by extending the work of de
Vega \Rc{\deVega}.  Without loss of generality we set $g=1$ and put
$S=-\ln \phi$ so that now
$$
J = {1\over\phi} \pmatrix{ 1 & \rhob \cr \rho & \phi^2 + \rho\rhob \cr}
\neqn\gaussforyang$$
and the self-duality equations reduce to Yangs equations
$$\eqalign{
\gMMb(\phi \delmu\delmub \phi -\delmu\phi \delmub\phi +
 \delmu\rho \delmub\rhob) &= 0  \cr
\gMMb(\phi \delmub \delmu\rho - 2 \delmu\rho \delmub\phi) = 0 \qquad
\gMMb( \phi \delmu \delmub\rhob - 2\delmu\phi \delmub\rhob) & = 0
\cr}\neqn\yangs$$
We shall immediately impose dimensional reduction, as in the last chapter, by
insisting that we have functions of $\eta_1$ and $\eta_2$ only.  Then one
trivial solution, extending a solution of de Vega, is that $\phi=1$,
$\rho=\rho_1(\eta_1)+\rho_2(\eta_2)$ and $\rhob=c(k_{12}\rho_1 - k_{21}
\rho_2) + b$ where b, c are arbitrary constants and $\rho_1$, $\rho_2$ are
arbitrary functions.  To obtain non-trivial solutions we  exploit the
bilinearity of the equations by making the Hirota style ansatz
$$
\phi={F\over\Delta} \qquad\qquad
\rho={N\over\Delta} \qquad\qquad  \rhob={\Nb\over\Delta}
\neqn\hirans$$
where
$$
\Delta=\Delta_0 + \Delta_1 e^{\eta_1} + \Delta_2 e^{\eta_2}
            +  \Delta_{12} e^{\eta_1 + \eta_2}
\qquad\qquad
F=F_0 +  F_1 e^{\eta_1} + F_2 e^{\eta_2} + F_{12} e^{\eta_1 + \eta_2}
$$
$$
N=N_0 +  N_1 e^{\eta_1} + N_2 e^{\eta_2} + N_{12} e^{\eta_1 + \eta_2}
\qquad\qquad
\Nb=  \Nb_0 +  \Nb_1 e^{\eta_1} + \Nb_2 e^{\eta_2} +
                                \Nb_{12} e^{\eta_1 + \eta_2}
\neqn\DFNNb$$

Substituting this ansatz into Yangs equations and equating to zero the
coefficients of the different powers of $e^{\eta_1}$ and $e^{\eta_2}$ gives a
set of 15 complicated equations in the 16 unknown coefficients.  Hence, the
system will indeed have solutions and de Vega produced a solution for the
special case in which $F_{12}=0$.  Here we point out that the equations
resulting from the above ansatz can be solved directly for rather general
conditions.  In order to do this we use a change of variables suggested by the
asymptotic properties of the ansatz.
Thus,
$$
\eta_2 \to -\infty \quad\Rightarrow\quad
 \phi(\eta_1,\eta_2) \to  \phi_-(\eta_1) = { F_0 + F_1 e^{\eta_1}
                            \over \Delta_0 + \Delta_1 e^{\eta_1}}
\neqn\phimin$$
and so the $\phi$ field is  a kink of height $f_1 = F_1/\Delta_1 -
F_0/\Delta_0$.  However it is $1/\phi$ and not $\phi$ that occurs in
$\chi$ and so we also look at
$$
 {1\over \phi_-(\eta_1)} = { \Delta_0 + \Delta_1 e^{\eta_1}
                            \over  F_0 + F_1 e^{\eta_1}}
\equiv \left({\Delta_1 \over F_1} - {\Delta_0 \over F_0}\right)
          g(\eta - \ln{F_0\over F_1}) + {\Delta_0\over F_0}
\neqn\invphimin$$
where $g(\eta):=(1+e^{\eta})^{-1}$ describes the shape (since it is
independent of the coefficients we only really get one type of wave in
this ansatz) and so $1/\phi_-$ is a kink with phase $\ln(F_0/F_1)$.
Similarly,
$$
\eta_2 \to \infty \quad\Rightarrow\quad
 \phi(\eta_1,\eta_2) \to  \phi_+(\eta_1) ={F_2 + F_{12} e^{\eta_1}
                            \over \Delta_2 + \Delta_{12} e^{\eta_1}}
\neqn\phimax$$
and so $1/\phi_+$ has phase $\ln(F_2/F_{12})$. Hence the phase shift between
these two limits is $\ln(F_0 F_{12}/F_1 F_2)$.  Furthermore the phase depends
only on the denominator and so only on $F$. In particular this means that
$1/\phi=\Delta/F$, $\rho/\phi=N/F$ and $\rhob/\phi=\Nb/F$ all suffer the same
phase shift.  From this it follows that it is natural to introduce the
variable $f_{12} := F_{12} F_0 - F_1 F_2$.  The phase shift is zero iff
$f_{12}=0$.  Thus we change to variables that describe the heights of the
kinks for $\phi$,$\rho$ and $\rhob$ by setting for i=1,2
$$
F_i = \Delta_i \left({F_0 \over \Delta_0} + f_i \right) \qquad
N_i = \Delta_i \left({N_0 \over \Delta_0} + n_i \right)  \qquad
\Nb_i = \Delta_i \left({\Nb_0 \over \Delta_0} + \nb_i \right)
\neqn\fnnbdef$$
and to variables suggested by the form of the phase shift for $1/\phi$ by
$$\eqalign{
F_{12} &= { F_1 F_2 + f_{12} \over F_0 } \qquad
\Delta_{12} = { \Delta_1 \Delta_2 + \delta_{12} \over \Delta_0 }   \cr
N_{12} &= { N_1 N_2 + n_{12} \over N_0 } \qquad
\Nb_{12} = { \Nb_1 \Nb_2 + \nb_{12} \over \Nb_0 }
\cr}\neqn\fiiidef$$
We assume that $\Delta_0 \Delta_1 \Delta_2 F_0 N_0 \Nb_0 \neq0 $, so that this
change of variables is non-singular.
If $k_{12}=-k_{21}$ or $k_{12}=0$ or $k_{21}=0$ then we have one of the cases
considered in the previous chapter, so we now also assume that $k_{12} k_{21}
(k_{12} +k_{21})\neq 0$.  Then the $e^{\eta_1 + \eta_2}$ terms of Yangs
equations immediately yield
$$\eqalign{
\delta_{12} &= {\Delta_0^2 \over F_0^2}  f_{12} +
  \Delta_1 \Delta_2 {\Delta_0^2 \over F_0^2}
\left( {n_1 k_{12} \nb_2 + n_2 k_{21} \nb_1 \over k_{12} + k_{21}} \right)
\cr
n_{12} &= {N_0^2 \over \Delta_0^2} \delta_{12}
+ 2 \Delta_1 \Delta_2 {N_0 \over F_0}
\left( {n_1 k_{12} f_2 + n_2 k_{21} f_1 \over  k_{12} + k_{21}} \right)
 - \Delta_1 \Delta_2 n_1 n_2
\cr
\nb_{12} &= {\Nb_0^2 \over \Delta_0^2} \delta_{12}
 + 2 \Delta_1 \Delta_2 {\Nb_0 \over F_0}
\left({f_1 k_{12} \nb_2 + f_2 k_{21} \nb_1 \over  k_{12} + k_{21}} \right)
 - \Delta_1 \Delta_2 \nb_1 \nb_2
\cr}\neqn\dnnbresul$$
The $\exp(\eta_1 +2 \eta_2)$ and $\exp(2\eta_1+\eta_2)$ terms
will give equations for $f_{12}$ and the full general solution then
depends on whether any of $f_1,\ldots,\nb_2$ are zero. However,
in order to obtain a solution with $f_{12}\neq0$ we find that we must impose
the constraints
$$
f_1^2 + n_1 \nb_1 =0  \qquad\qquad
f_2^2  + n_2 \nb_2 =0
\neqn\constraints$$
and we then obtain a solution of Yangs equations as long as
$$
f_{12} = - \Delta_1 \Delta_2 (2 f_1 f_2 + n_2 \nb_1 + n_1 \nb_2)
   { k_{12} k_{21} \over (k_{12} + k_{21})^2}
\neqn\fiiires$$
In this generic solution, besides the $k_{ij}$ we have the 12 parameters
$\Delta_0$, $\Delta_1$, $\Delta_2$, $F_0$, $N_0$, $\Nb_0$, $f_1$, $f_2$,
$n_1$, $n_2$, $\nb_1$ and $\nb_2$ subject to the two constraints
\E\constraints.  However we can trivially set $\Delta_0=1$, and by shifting
the coordinate origin we could set $\Delta_1=\Delta_2=1$ and so there are
really only 12-2-3=7 true parameters to this general solution.  Also the
overall scale leaves $A_{YM}$ unchanged (see expressions for A given by de
Vega) and so might not be considered a true parameter.

With this solution it turns out that $(\phi^2 + \rho\rhob)/\phi$ is also of
the form of the Hirota ansatz and so effectively we have simply imposed such
an ansatz directly on the components of J.

In order to interpret the solution we can look at the behaviour of the stress
energy tensor of \E\TdV.  We find that for $\eta_2 \to -\infty$ we have that T
becomes proportional to the constraints \E\constraints and so for this
solution generically T will become zero at $\eta_i=\pm\infty$. Hence in the
above solution to Yangs equations the stress energy generically vanishes at
long distance from the region where the kinks are colliding.  In this respect
is is more like an instanton solution of the underlying 2 dimensional chiral
model.  The particular solution that de Vega found is the special case
$F_{12}=0$ and this is a degenerate case in which the stress tensor spreads
out infinitely in one particular direction.

If we impose reality conditions to try to get an SU(2) solution then we
inevitably find we are forced to trivialise the solution.  For example, in the
$\C^{1,1}$ case we must impose $\rho^*=\rhob$ and $\phi^*=\phi$; whence also
$n_i^*=\nb_i$ and $f_i^*=f_i$ and then \E\constraints\ forces
$f_i=n_i=\nb_i=0$ and in particular $f_{12}=0$. The reality conditions are
homogeneous and so it is not unreasonable to have tried the Hirota ansatz in
this case.  There is no problem if we want SL(2,R) or SU(1,1) as we simply
take a real solution or solve
\E\constraints with $n_i^*=-\nb_i$.  Perhaps one point worth mentioning with
respect to SU(2) solutions is that in a (2,2) signature the B\"acklund
transformations in \Rc{\Corriganetal}\ now preserve the reality conditions
and so take SU(2) solutions to SU(2) solutions.  They no longer alternate
between SU(1,1) and SU(2) solutions as happens in a (4,0) signature.

An obvious question is whether this can be repeated for other gauge groups.
The main step is the use of an Hirota ansatz and this will only be useful for
homogeneous equations.  Since \E\pohl are homogeneous in the matrix elements of
J for any G then we expect that GL(n) could be treated in this way.  However,
as we just saw, this will not usually work for subgroups because the extra
conditions are usually not homogeneous and so not suitable for the Hirota
ansatz.  Note these solutions are just scattering in a general chiral model,
which is integrable, and so in principle all solutions are known, but in
practise are not trivial.

\chapter{Conclusion and open questions}

In this paper we briefly reviewed the kinematics in (complexified) space-time
with a (2,2) signature. We found a simple way to obtain the three-nulls
identity that is vital to calculations of on-shell Feynman diagrams.  We then
briefly covered the J formulation of SDYM, its perturbation theory and
dimensional reduction to a two dimensional chiral model with Wess-Zumino term.

As pointed out in \Rc{\OoguriVi}\ a dominant feature of the perturbation
theory is that the on-shell connected amplitudes vanish for four or more
external legs.  This leaves only three particle scattering and then the
momentum dependence of the vertices has the implication that two particles
(physical states) can only interact if their momenta are in the same self-dual
null plane.  This only applies to the S-matrix but not to Greens functions but
does seem to imply that the effective dimensionality has been reduced by one.
It would be interesting to classify all possible actions for which the
S-matrix has the same properties.  For example we might observe that the
action \E\newpohlaction\ is a non-symmetric non-linear sigma model, and so
instead we could generalise to
$$
L = G_{ab}[\phi] \delmu \phi^a \delmub \phi^b   \gMMb
\neqn\sigmodel$$
Enforcing the vanishing of the appropriate Feynman diagrams will presumably
then have some geometric meaning for the non-symmetric metric on the space
parametrised by the $\phi^a$.  Presumably including fermions will be related
to SDYM with supergroups or to some supersymmetric extension of the
self-duality relation: It should be noted that a tree level diagram with all
bosonic external legs cannot have any internal fermion lines and so the
bosonic sector of any such general theory should already possess the property
that only the three point function is non-trivial.  The S-matrix predicted by
the perturbation theory is very simple, and if this structure is still present
non-perturbatively then it would be very interesting to find the consequences
for the S-matrix (compare factorisation of the S-matrix in two dimensions).

A classical equivalent of particle momenta being in the same self-dual plane
is that the fields depend only on the two coordinates of the plane (or more
generally to be non-linear superpositions of such fields on different planes)
In this case SDYM reduces to the equation of motion from a chiral model in
(1,1) spacetime with the Wess-Zumino term only.  Such a ``pure WZ'' model is
classically exactly solvable and topological.  It is also not an evolution
equation and so this leads us to question the validity of the perturbation
theory prediction for the scattering in the theory.  (This also means that
SDYM is on the edge of being topological \Rc{\OoguriVii}\ and so we might
expect some of the physical observables to be non-local and measured by
appropriate Wilson lines.)

In order to see whether these properties hold non-perturbatively in the
coupling constant (but still at tree level) we were lead to consider exact
classical solutions of SDYM.  In particular there are certainly some exact
solutions which correspond to 2 to 2 scattering with no momentum exchange but
a non-zero phase shift.  For example the theory can be reduced to the KdV
equation or a (1,1) chiral model.  In this case the momenta would not
necessarily lie on the same self-dual plane and so the factorisation
considered above would be destroyed (and the perturbation theory would have
been very misleading).  To clarify this we extended some work of de Vega
\Rc{\deVega}\ on the scattering of plane waves in SL(2,C).

Using the Hirota ansatz and Gaussian decomposition of the SL(2,C) field we
were able to solve the SDYM equations for the case of a generalised plane
wave.  The solution described the collision of kinks in the J field, and
consisted of non-trivial scattering in the sense that the kinks suffered a
non-zero phase shift from the collision.  However the restrictions necessary
to obtain such a solution were also precisely sufficient to force the in and
out waves to have vanishing stress-energy tensor and Yang-Mills field
strengths.  That is, the stress tensor (meaning the one from the action for
Yangs equations not the one from the Yang-Mills action) is non-zero only in
the region where the kinks are colliding.  On the positive side, this does not
seem to correspond to 2 to 2 scattering of physical states and so will not
affect the above S-matrix properties.  The negative side is that there is
still the possibility that there is no scattering of physical states.

In our SL(2,C) solution we have a two dimensional surface of non-zero energy
embedded in the spacetime which is reminiscent of a string theory again.  It
would be interesting to investigate whether such solutions have a stringlike
behaviour in which strings beget strings along the lines suggested in
\Rc{\Green}.
\ack  {I would like to thank C. Devchand for useful discussions.
I am also grateful to the Royal Society of Great Britain and the  Swiss
National Science Foundation for financial support.}
\refout
\bye